\documentclass[10pt]{article}
\usepackage{amssymb}
\usepackage{amsbsy}
\usepackage{amsmath}
\usepackage{amsfonts}
\usepackage{verbatim}
\usepackage[dvips]{graphicx}
\usepackage{mathenv}
\begin{document}
\begin{flushright}
 BI-TP 2005/23\\
LPT-Orsay 05-53
\end{flushright}
\vspace{1.0cm} {\begin{center}

\huge {\bf  Has saturation physics been observed in deuteron-gold
collisions at RHIC ?}\end{center}}

\vspace{1.0cm}

\begin{center}
{Rudolf Baier $^{a}$, Yacine Mehtar-Tani $^{b,}$\footnote{E-mail
address: mehtar@th.u-psud.fr}, Dominique Schiff $^{b}$}

\vspace{0.5cm} $^a${\it Fakult\"at f\"ur Physik, Universit\"at
Bielefeld,\\D-33501,
Bielefeld, Germany } \\
$^b$ {\it Laboratoire de Physique Th\'eorique, Universit\'e de
Paris XI, \\B\^atiment 210,
91405 Orsay Cedex, France} \\
\end{center}\vspace{0.5cm}

\begin{abstract}
  In the framework of the recently proposed saturation picture, we examine in a
  systematic way whether the nuclear modification factor measured for d-Au
  collisions at RHIC may be simply explained. The Cronin peak which is obtained
  at mid-rapidity around $k_{\bot}\simeq 3$ GeV may be reproduced at the proper
  height only by boosting the saturation momentum by an additional nuclear
  component as already shown in the literature. In this respect, mid-rapidity
  RHIC data cannot necessarily be seen as a probe of the saturation picture. The large rapidity ($\eta\simeq
  3$) region allows us to test the shape of the unintegrated gluon distribution
  in the nucleus, investigating various parameterizations inspired by large
  rapidity solutions (of the BFKL and) of the Balitsky-Kovchegov (BK) equation. A satisfactory
  description of $R_{CP}$ at RHIC is
  obtained in the BK picture.
\end{abstract}
  \section{Introduction}

  Testing the saturation (Color Glass Condensate) picture
  \cite{JKrev}, for the
  initial state of deuteron-gold (dA) collisions against RHIC data has been a
  subject of interest for some time. Two salient features have been observed
  \cite{BRAHMS1,exp} concerning the behavior of the nuclear modification factor $R_{dA}$.
  At mid-rapidity, the Cronin peak height depends on the centrality of the
  collision. At large rapidity, the suppression predicted by quantum evolution is
  observed and is bigger for smaller centralities. A number of papers have recently discussed the description of the Cronin enhancement \cite{BKW,AG,KKT1,KKT2,JMG,IIT} and the effect of small $x$ \cite{KKT1,KT,AAKSW,JJM,KLM,BGV1}. In this work, we examine in a
  systematic way, how and if saturation and quantum evolution provide a
  reasonable quantitative agreement with data \cite{BRAHMS1}.\\
  The time is indeed appropriate to assess the predictability of
  the saturation (CGC) picture. This endeavor  has a mitigated
  conclusion: as it turns, unavoidably, the saturation scale
  introduced in the theory, does not have the proper size to
  explain RHIC data at mid-rapidity. This conclusion is similar to the one stated in \cite{AG}. On the other hand, quantum
  evolution as described by the theory gives the proper
  suppression of the nuclear modification factor above the
  saturation scale.

    In section 2.1, we calculate the hadron production cross section in dA at
  mid-rapidity using the semi-classical approach and show the prediction
  for the minimum-bias nuclear modification factor
  $R_{dA}$ (and $R_{CP}$ for central versus peripheral collisions dA collisions)
  in relation with other previous studies and comparing with data.\\
   We then discuss, in section 2.2, quantum evolution. We first derive the expression for the
  cross-section at leading log accuracy, including both gluon and quark
  distributions within the deuteron. This expression is identical to eq. (22) in
  \cite{dum}. We then present various parameterizations of the unintegrated gluon
  distribution in the nucleus, inspired by large rapidity solutions of the BK
  equation \cite{BK}
  and show the comparison with representative data \cite{BRAHMS1}.
  In section 3, the conclusion and outlook are given.

\section{Hadron production in dA}\label{sec1}

The nuclear modification factor $R_{dA}$ and the
  $R_{CP}$(Central/Peripheral collisions)
ratio are defined as
\begin{equation}\label{RpA}
R_{dA}=\frac{1}{N_{coll}}\frac{\frac{dN^{dA\rightarrow hX}}{d\eta
d^{2}{\bf k} }}{\frac{dN^{pp\rightarrow hX}}{d\eta d^{2}{\bf k}
}},
\end{equation}
\begin{equation}
R_{CP}=\frac{N^{P}_{coll}\frac{dN^{dA\rightarrow hX}}{d\eta
d^{2}{\bf k} }\vert_{C}}{N^{C}_{coll}\frac{dN^{dA\rightarrow
hX}}{d\eta d^{2}{\bf k}}\vert_{P}}.
\end{equation}
${\bf k}$ and $\eta$ are respectively the transverse momentum and
the pseudo-rapidity of the observed hadron. $N_{coll}$ is the
number of collisions in dA, it is roughly twice the number of
collisions in pA(proton-Gold). The centrality dependence of
$R_{dA}$ is related to the dependence of $N^{dA\rightarrow
hX}=d\sigma^{dA\rightarrow hX}/d^{2}{\bf b}$ and $N_{coll}({\bf
b})$ on the impact parameter of the collision. In this paper, we
address the predictions of the Color Glass Condensate for these
ratios. We always assume that cross-sections depend on the impact
parameter only through the number of participants which is
proportional to the saturation scale
\begin{equation}
Q_ {sA}^2({\bf b})\simeq Q_ {sA}^2(0)N _{part.Au}({\bf b})/N
_{part.Au}(0),
\end{equation}
where $N_{part.Au}$ is the number of participants in the gold
nucleus in d-Au collisions. This is coherent with the assumption
that $Q_ {sA}^2({\bf b}) \simeq (N _{part.Au}({\bf b})/2)Q_
{sp}^2$ such that $ Q_ {sA}^2({\bf b}=0)\simeq A^{1/3} Q_ {sp}^2$
\cite{KLN}.
 We use Table 2 in \cite{BRAHMS1} which gives the number
of participants $N_{part}$ and the number of collisions $N_{coll}$
for several centralities.
\subsection{Semi-classical approach}
\label{sec11}

 We first deal with gluon production at
mid-rapidity for which different approaches have been proposed.
The inclusive cross section has been calculated in \cite{KovMul}
in a quasi classical approach of multiple rescattering inside the
nucleus, see also \cite{KKT1,KT}. Further confirmation has been
given in \cite{BGV1,DuMcl}. The
  inclusive cross section, for a gluon with transverse momentum ${\bf k}$
and
  rapidity $\eta=0$, is written as

\begin{equation}\label{KKT}
\frac{d\sigma^{dA\rightarrow gX}}{d\eta d^{2}{\bf k} d^{2}{\bf
b}}=\frac{C_{F}}{\alpha_{s}\pi(2\pi)^{3}}\frac{1}{{\bf k}^{2}}\int
d^{2}{\bf B}\int d^{2}{\bf z} \nabla^{2}_{{\bf z}}n_{G}({\bf
z},{\bf b}-{\bf B})\nabla^{2}_{{\bf z}}N_{G}({\bf z},{\bf
b})e^{-i{\bf k}.{\bf z}},
\end{equation}
where $N_{G}({\bf z},{\bf b})$ is the forward scattering amplitude
of a gluon dipole off the nucleus, it contains all higher twists
in the semi classical approximation and $n_{G}({\bf z},{\bf
b}-{\bf B})$ is the forward scattering amplitude of a gluon dipole
off the deuteron at leading twist approximation. ${\bf B}$ and
${\bf b}$ are the impact parameters of the deuteron and the gluon
with respect to the center of the nucleus. At mid-rapidity and
RHIC energies, it is legitimate to neglect quantum evolution.

This approach provides a description of $N_{G}$ at all orders in
terms of the saturation scale $Q^{2}_{sA}=Q^{2}_{s}\varpropto
A^{1/3}\Lambda^{2}_{QCD}$. The CGC approach yields a
Glauber-Mueller form for the dipole forward scattering amplitude
\begin{equation}\label{NG}
N_{G}({\bf z},{\bf b})=1-{\bf exp}({-\frac{1}{8}{\bf
z}^2Q^{2}_{s}({\bf b})\ln\frac{1}{{\bf z}^2\Lambda^2}}),
\end{equation}
where the saturation scale is given by  \cite{KovMul}
\begin{equation}\label{QS}
Q^{2}_{s}({\bf b})\ln\frac{1}{{\bf
z}^2\Lambda^2}=\frac{4\pi^{2}\alpha_{s}}{C_{F}}\rho T({\bf
b})xG(x,1/{\bf z}^{2}).
\end{equation}
$\rho$ is the nuclear density in the nucleus  and $T({\bf
b})=2\sqrt{ R^2-{\bf b}^2}$ the nuclear profile function of a
spherical nucleus of radius $R$. $\Lambda$ is an infrared cut-off
of order $\Lambda_{QCD}$. The cross section can be rewritten in
the following simpler form
\begin{equation}\label{cross}
\frac{d\sigma^{dA\rightarrow gX}}{d\eta d^{2}{\bf k} d^{2}{\bf
b}}=\frac{C_{F}\alpha_{s}}{\pi^{2}}\frac{2}{{\bf
k}^{2}}\int_{0}^{~1/\Lambda}du\ln\frac{1}{u\Lambda}\partial_{u}[u\partial_{u}N_{G}(u,{\bf
b })]J_{0}(\vert {\bf k}\vert u),
\end{equation}
where $u=\vert {\bf z}\vert$.\\
 \subsubsection{A model for minimum bias collisions}

   One defines the minimum bias cross-section as the average over the impact parameter of the collision, it may be written as
 \begin{equation}
 \frac{dN^{min.bias}}{d\eta d^{2}{\bf k}}=\langle \frac{d\sigma}{d\eta d^{2}{\bf
 k}d^{2}{\bf b}}\rangle=\frac{1}{S_{A}} \int d^{2}{\bf b}\frac{d\sigma}{d\eta d^{2}{\bf
 k}d^{2}{\bf b}}.
 \end{equation}
 $S_A=\pi R^2$ is the transverse area of the nucleus. All the centrality dependence in the cross-section is contained in $N_G({\bf z},{\bf b})$ as given by (\ref{NG}) and (\ref{QS}).
 One can perform the ${\bf b}$ integral for $N_G$ yielding
\begin{equation}
 \langle N_{G}({\bf z},Q_{s})\rangle=1+\frac{128}{{\bf z}^4 Q_{sC}^4 \ln^2
 (1/{\bf z^2}\Lambda^2)}[(1+\frac{1}{8}{\bf z}^2Q_{sC}^2 \ln
 (1/{\bf z}^2\Lambda^2))\exp\{-\frac{1}{8}{\bf z}^2 Q_{sC}^2 \ln
 (1/{\bf z}^2\Lambda^2)\}-1],
 \end{equation}
 where $Q_{sC}^2=Q^2_s ({\bf b=0})$. The corresponding result for $R_{dA}$ is shown in Fig.
 \ref{fig1} (a) taking $Q_{sC}^2=2$ GeV$^2$ and $\Lambda=0.2$ GeV. The proton-proton cross-section is  calculated by using the Glauber-Mueller formula (\ref{NG}) with $Q_{sp}$. $R_{dA}$ shows a Cronin peak for $k_{\bot}$
in the range of $Q_{s}$. We have used a prescription  such that
the region  $z \sim 1/\Lambda$ does not affect the integral. For
that purpose, we make the replacement $\ln\frac{1}{{\bf
z}^2\Lambda^2} \rightarrow \ln(\frac{1}{{\bf
z}^2\Lambda^2}+a^{2})$ choosing $a=3$ \cite{BKW}.
 A good approximation of the integral
 over ${\bf b}$ is to choose for $Q_{s}$ the
 average value $\langle Q^{2}_{s}({\bf
 b})\rangle\equiv Q^{2}_{s.min-bias}=(2/3)Q_{sC}^2$, in which case
\begin{equation}
\label{Rb}
 \langle \frac{d\sigma}{d\eta d^{2}{\bf
 k}d^{2}{\bf b}}(Q^{2}_{s}({\bf b}))\rangle\simeq \frac{d\sigma}{d\eta d^{2}{\bf
 k}d^{2}{\bf b}}(\langle Q^{2}_{s}({\bf b})\rangle).
 \end{equation}
 Actually, (\ref{Rb}) turns into an equality in the high $k_{\bot}$ region
 and is quite good in the region of the Cronin peak. In Fig. \ref{fig1}(b) we see that the error is
 maximum when ${\bf k}\lesssim Q_{s}$, reaching $10 \%
 $. For a
 cylindrical nucleus there is no ${\bf b}$ dependence in the cross
 section  and (\ref{Rb}) turns into an equality.
 This tells us that the physics is the same whatever the
 geometry of the nucleus \cite{KKT1}.
 \subsubsection{Effects of fragmentation on the Cronin peak}

  To get the hadron  cross-section we still have to convolute
(\ref{cross}) with the proper fragmentation functions
  \begin{equation}\label{crossFF}
\frac{d\sigma^{dA\rightarrow hX}}{d\eta d^{2}{\bf k} d^{2}{\bf b}}
= \int \frac{dz}{z^{2}}D_{g}^{h}(z,Q_f
^2)\frac{d\sigma^{dA\rightarrow gX}}{d\eta d^{2}{\bf q} d^{2}{\bf
b}}({\bf q}={\bf k}/z),
\end{equation}

\begin{figure}[hbtp]
\begin{tabular}{c  c }

\qquad\includegraphics[width=6cm]{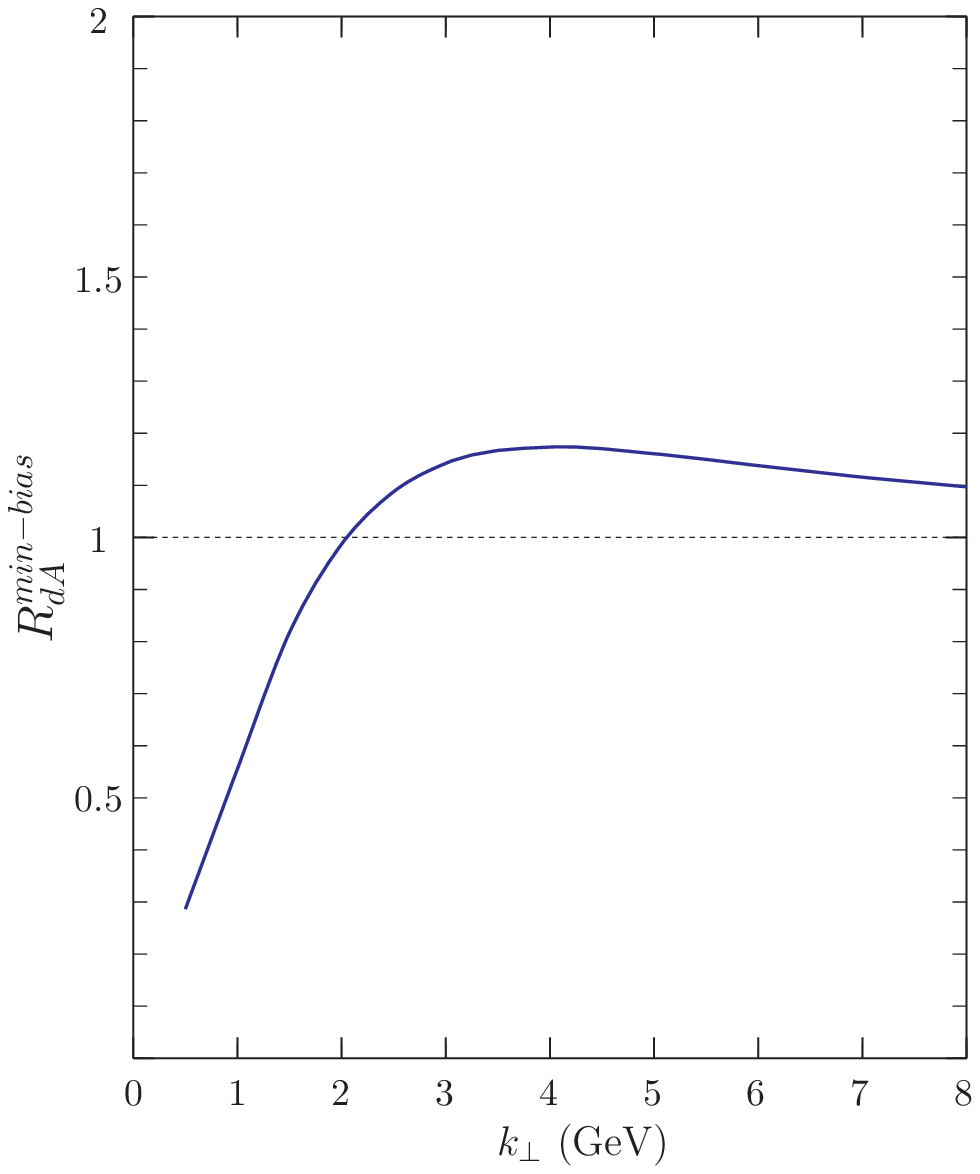} &\qquad   \includegraphics[width=6cm]{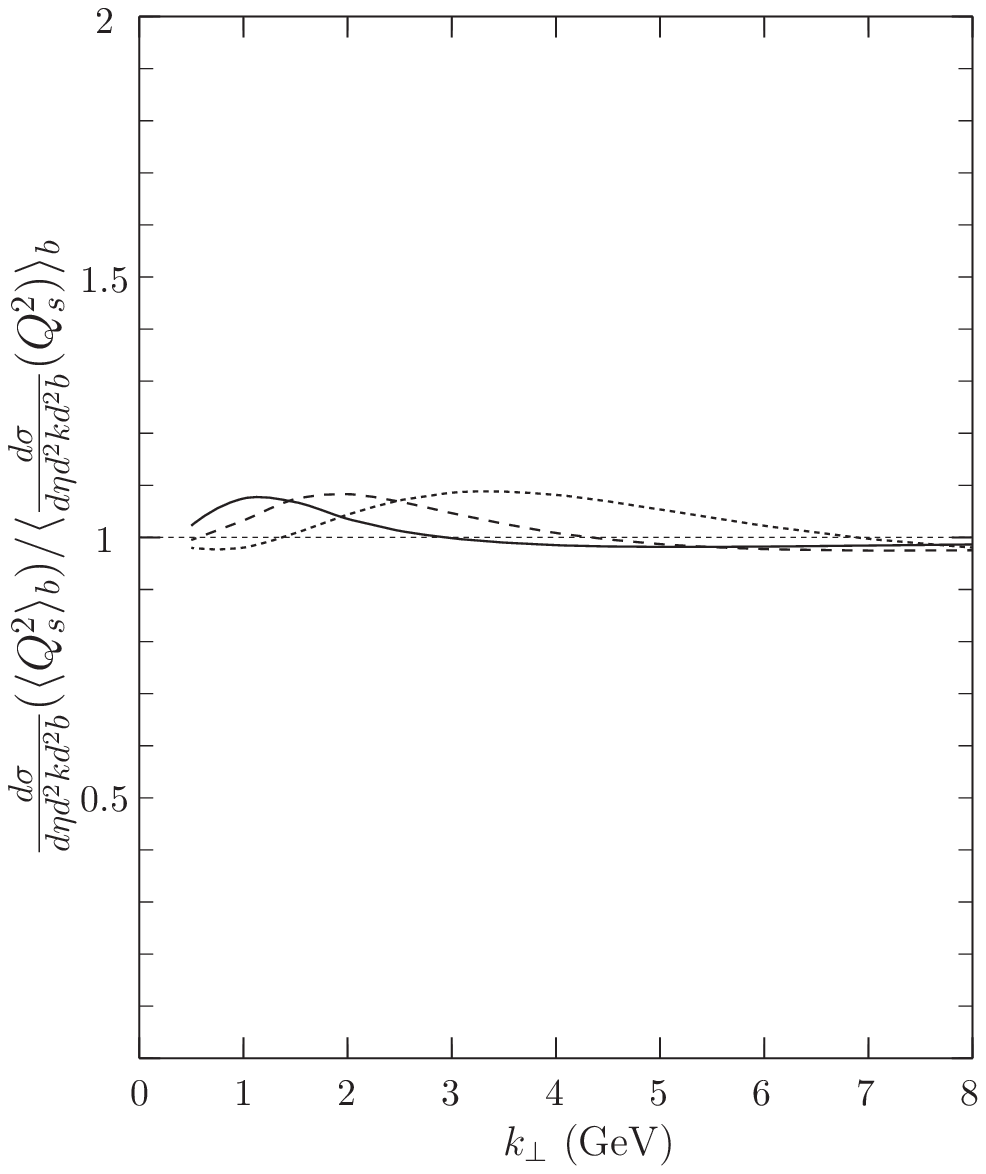} \\
      \qquad (a)& \qquad(b)
\end{tabular} \caption{(a) :
The nuclear modification factor $R_{dA}$ without fragmentation
functions (gluon production) for $Q^2_{sC}=2$ GeV$^2$. (b) :
Comparison between the average full min-bias calculation and the
calculation  with average min-bias $Q_s$ for $Q^2_{sC}=2, 5$ and
$9$ GeV$^2$ (Full, dashed and dotted).} \label{fig1}
\end{figure}
\begin{figure}[hbtp]
\begin{tabular}{c  c }

\qquad\includegraphics[width=6cm]{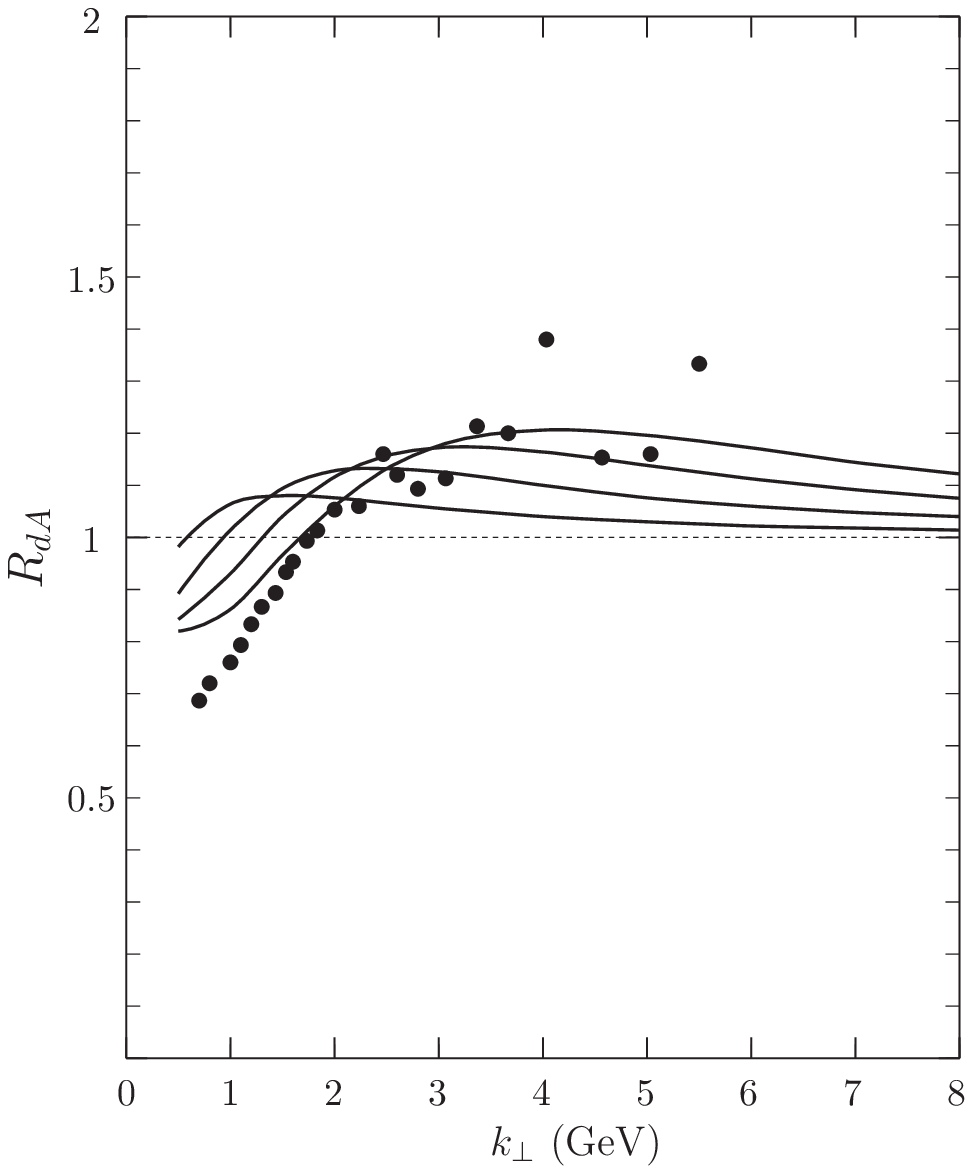} &\qquad   \includegraphics[width=6cm]{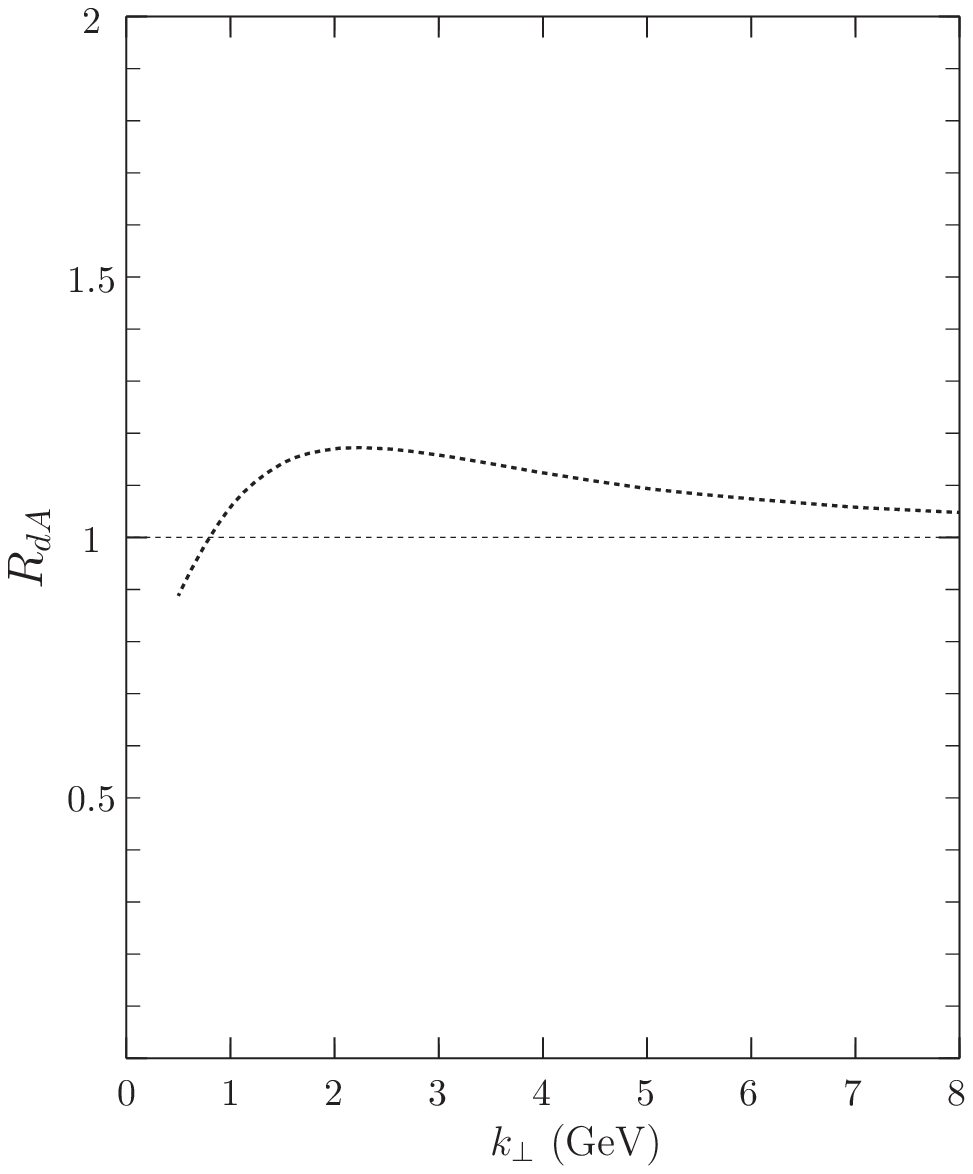} \\
      \qquad (a)& \qquad(b)
\end{tabular}\caption{(a) :
Minimum bias $R_{dA}$ for charged hadron production for
$Q^2_{s.min-bias}=1.3, 3.33, 6$ and $9$ GeV$^2$ (lowest to highest
curve). (b) : Minimum bias $R_{dA}$ for neutral pions production
for $Q^2_{s.min-bias}=9$ GeV$^2$. The points are representative
data taken from \cite{BRAHMS1}.} \label{fig2}
\end{figure}

where $Q_f$ is a large scale  of order $k_{\bot}$ taken between 2
and 8 GeV. We use the fragmentation functions of \cite{kkp}. In
Fig.\ref{fig2}(a) we present the result for $R_{dA}$.
 
\begin{figure}[h]
\centering
\includegraphics[width=8cm]{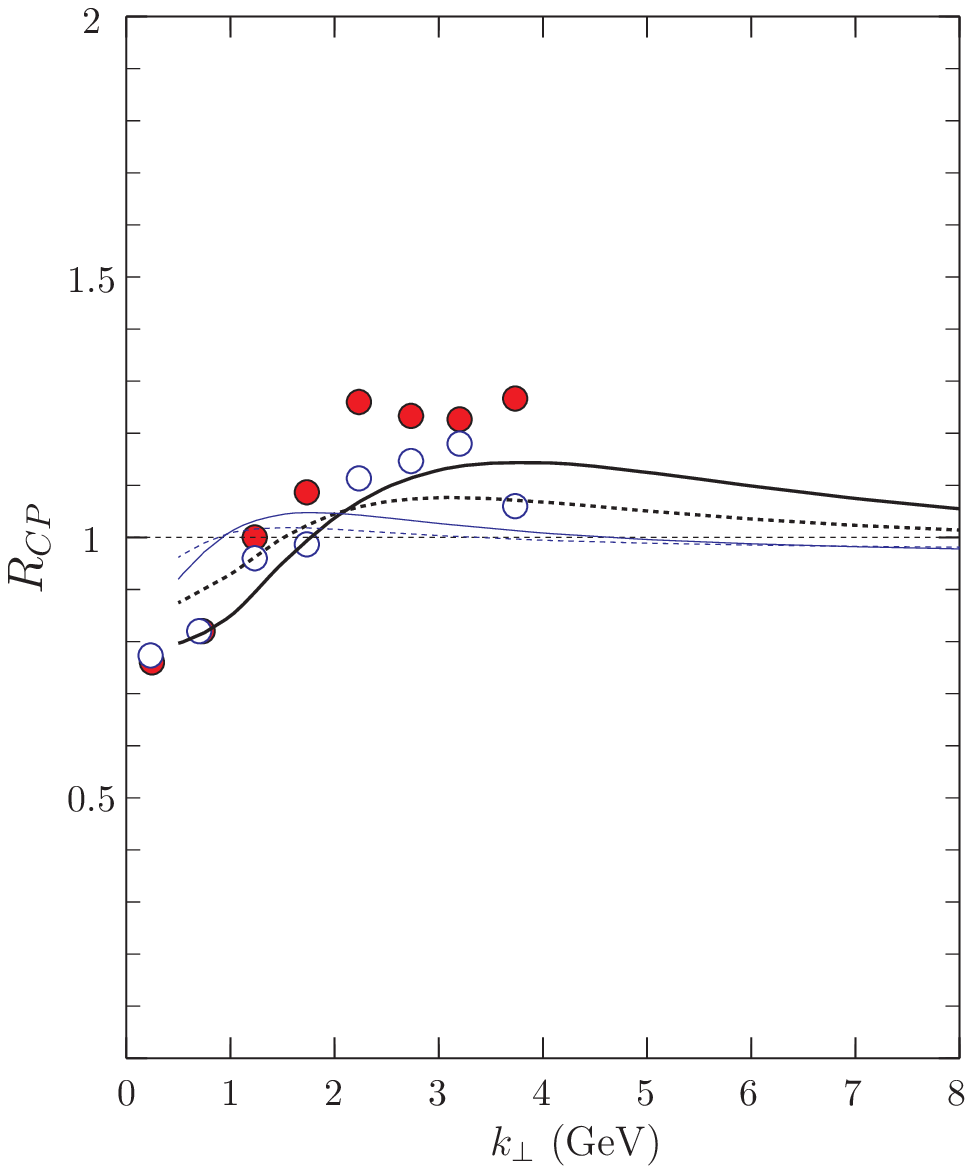}

\caption{$R_{CP}$ for $Q^2_{sC}=9$ GeV$^2$ (thick lines) and
$Q^2_{sC}=2$ GeV$^2$ (thin lines). Full lines correspond to
central over peripheral collisions (full experimental dots).
Dashed lines correspond to semi-central over peripheral collisions
(empty experimental dots). Data from \cite{BRAHMS1}.}\label{fig3}
\end{figure}

  The fragmentation functions induce a flattening of the Cronin
peak toward 1, with a shift toward the small $k_{\bot}$'s. This
feature can be understood by comparing $R^g_{dA}$ without F.F's
(fragmentation functions) and $R^h_{dA}$ with F.F's, order by
order in powers of $Q_{s}^2/{\bf k}^{2}$ (or the number of
participants). Above the saturation scale, the following expansion
has a meaning
\begin{equation}
R^g_{dA}=1+R_{dA}^{(1)g}+...
\end{equation}
In the Leading Log approximation with respect to
$\ln(k_{\bot}/\Lambda)$, using (\ref{crossFF}), we approximate the
nuclear modification factor by
\begin{equation}
R_{dA}^h=1+\frac{\langle z^4 \rangle}{\langle z^2
\rangle}R_{dA}^{(1)g}+...
\end{equation}
where we have defined for $n\geqslant 2$
\begin{equation}
\langle z^n \rangle=\int_{z_0}^1 dz D(z,Q_f^2) z^n/\int_{z_0}^1 dz
D(z,Q_f^2).
\end{equation}
Since $z<1$ we always have $\langle z^n \rangle/\langle z^2
\rangle <1$. The rescattering terms are then less important and
$R_{dA}^h$ gets closer to 1. This is shown in Fig. \ref{fig2} (a) :
comparing with  Fig. \ref{fig1} (a) we see the dramatic effect of
the fragmentation functions for $Q_{s.min-bias}^{2}=1.3$ GeV$^2$.
To get agreement with RHIC data we have to increase the value of
the saturation scale $Q_s$ and take $Q_{s.min-bias}^{2}\gtrsim 6$
GeV$^2$. This feature has already been mentioned in \cite{KKT2}.
In fact, $Q_s^2$ as defined in (\ref{QS}) is at most of the order
of $2$ GeV$^2$ for ${\bf b}=0$. The authors of \cite{KKT2} propose
to enhance it by adding an additional momentum due to
"non-perturbative" nuclear effects :
\begin{equation}\label{ansatz}
Q_s^2\rightarrow Q_s^2+\kappa^2 A^{1/3},
\end{equation}
with $\kappa^2=1$ GeV$^2$. This amounts to boosting $Q_s^2({\bf
b}=0)$ to $\sim9$ GeV$^2$. In fact, as shown in Fig. \ref{fig2} (a), we need even a larger saturation scale. On the other hand,
the way $R_{dA}$ is normalized (by estimating the proton-proton
cross-section using (\ref{NG})) is not fully convincing. The ratio
$R_{CP}$ does not suffer from the same uncertainty. It is shown in
Fig. \ref{fig3} for $Q_{sC}^2=9$ GeV$^2$ and $Q_{sC}^2=2$ GeV$^2$
for central and semi-central collisions. Taking into account the
experimental error-bars, the large value for $Q_{sC}$ is
definitely preferred and in agreement
with data.\\

We should at this point remark that saturation physics in the
present stage is in the situation where the leading order
perturbative QCD description was, concerning large $k_{\bot}$
spectra in hadron-hadron collisions. The agreement with data could
only be obtained by implementing an intrinsic non-perturbative
transverse momentum for partons inside the hadron. The present
state seems to be that, at next-to-leading order, the perturbative
theory becomes predictive \cite{AF,Av} : a detailed NLO comparison with RHIC data for $pp\rightarrow\pi^0X$ is presented in \cite{Adler}. In this respect and in the
leading order perturbation QCD framework, the authors of ref.
\cite{AG} have used a traditional Glauber-Eikonal approach of
sequential multiple partonic scatterings with the implementation
of a large intrinsic $k_{\bot}$ in parton distribution functions
and they have obtained a good agreement with data for $\pi_0$
production in dA collisions.\\

 Staying in the saturation physics
framework, we may nonetheless try to go beyond the ad-hoc ansatz
(\ref{ansatz}) and modify the saturation picture, which is
exclusively based on hard multiple scatterings, by adding
non-perturbative scatterings, when a parton (gluon) is passing
through a nucleus. Following the Moli\`ere scattering theory
\cite{Bethe} extended to QCD \cite{BDMPS}, we define the
probability distribution for the scattered parton by
\begin{equation}\label{Nphi}
V({\bf k})=\frac{1}{\sigma}\frac{d\sigma}{d^2{\bf
k}}=\frac{1}{\pi}\{\frac{c}{\langle {\bf k}^2\rangle
}e^{-\frac{{\bf k}^2}{\langle {\bf k}^2\rangle
}}+\frac{(1-c)\Lambda^2}{({\bf k}^2+\Lambda^2)^2}\},
\end{equation}
with $\int d^2{\bf k} V({\bf k})=1$.\\

A soft, non-perturbative, gaussian contribution is added to the
hard screened (by mass $\Lambda$) gluon exchange term. Solving the
kinetic master equation for the survival probability of the
propagating parton, an effective scale (up to logarithm) is
derived,
\begin{equation}
Q_s^2\rightarrow (c\frac{\langle {\bf
k}^2\rangle}{\Lambda^2}+(1-c))Q_s^2.
\end{equation}
Different from (\ref{ansatz}) the non-perturbative part is added
but
weighted by a factor $c<1$.\\

In order to obtain an effective $Q_s^2\approx O(10\text{ GeV}^2)$
one has to add a significant non-perturbative part, e.g. for
$\langle {\bf k}^2\rangle\simeq 0.5$ GeV$^2$, $\Lambda=0.2$ GeV
and $c\simeq0.3$ indeed
\begin{equation}
Q_s^2\approx2\text{ GeV}^2\rightarrow Q_s^2\approx 9\text{ GeV}^2.
\end{equation}
The underlying picture of dominating soft parton interactions with
nucleons has strong implications, especially for the width of jet
broadening, i.e. the resulting transport coefficient for cold
matter $\hat{q}\simeq Q_s^2/(2R)$ becomes also rather large,
namely $\hat{q}\simeq 0.8$ GeV$^2$/fm (for gold)! But such a
strong jet broadening has not been observed so far \cite{arleo}.
\subsection{Forward rapidity and quantum evolution}
\label{sec12}
\subsubsection{Hadron production cross-section}

  As shown in \cite{KKT1,KT}, it is possible to rewrite eq. (\ref{KKT})
under a $k_{\bot}$-factorized form \cite{KovMul}, which is then generalized to
include the rapidity dependence. At leading twist, on defines the unintegrated gluon
distribution for the nucleus and the deuteron, respectively, as
\begin{equation}\label{Nphi}
\varphi_{A}(Y_{A},{\bf k},{\bf
b})=\frac{C_{F}}{\alpha_{s}(2\pi)^{3}}\int d^{2}{\bf
z}\nabla^{2}_{{\bf z}}N_{G}(Y_{A},{\bf z},{\bf b})e^{-i{\bf
k}.{\bf z}},
\end{equation}
and
\begin{equation}
\varphi_{p}(Y_{d},{\bf k},{\bf b}-{\bf
B})=\frac{C_{F}}{\alpha_{s}(2\pi)^{3}}\int d^{2}{\bf
z}\nabla^{2}_{{\bf z}}n_{G}(Y_{d},{\bf z},{\bf b}-{\bf
B})e^{-i{\bf k}.{\bf z}},
\end{equation}
with $Y_{A}=\ln(1/x_{A})=Y+\eta$ and $Y_{d}=\ln(1/x_{d})=Y-\eta$
the rapidities of the gluons merging respectively from the nucleus
and the deuteron and carrying the light cone momentum fractions
$x_{A}$ and $x_{d}$; $\eta$ is the rapidity of the produced gluon
measured in the forward deuteron direction and
$Y=\ln(\sqrt{s}/k_{\bot})$. One obtains the expression for the
cross-section as
\begin{equation}\label{kt}
\frac{d\sigma^{dA\rightarrow gX}}{d\eta d^{2}{\bf k} d^{2}{\bf
b}}=\frac{2\alpha_{s}}{ C_{F}}\frac{1}{{\bf k}^{2}}\int d^{2}{\bf
B}\int d^{2}{\bf q }\varphi_{d}({\bf q}-{\bf k},Y-\eta,{\bf
b-B})\varphi_{A}({\bf q},Y+\eta,{\bf b}).
\end{equation}
We will focus our analysis on the large forward rapidity region for the gluon where we can neglect the emission of
additional gluons in the wave function of the proton. Actually, in
this region the biggest value of $Y_d$ is reached at $\eta=0$
yielding $Y_{d}=Y$ which is not indeed large enough at the
energies of RHIC for the evolution to take place. However, the
opposite happens in the nucleus where $Y_{A}$ increases with
$\eta$ and we thus expect to probe the small $x$
 regime of the nucleus wave function \footnote{A calculation of $R_{dA}$ has been recently performed \cite{Strik} in the framework of linearly factorized pQCD at NLO, taking into account nuclear leading twist shadowing for the partonic distributions in the nucleus. The observed suppression cannot be explained in this context.}. We have, taking $\varphi_{d}\simeq
2\varphi_{p} $ at leading twist without quantum evolution
\cite{KovMul},
\begin{equation}
\int d^{2}{\bf b} \varphi_{d}({\bf k},Y-\eta\simeq 0,{\bf
b})=\frac{\alpha_{s}C_{F}}{\pi}\frac{2}{{\bf k}^{2}}.
\end{equation}
This allows us, assuming $\varphi_{A}$ smooth enough, to
approximate (\ref{kt}) at leading log accuracy, when $\vert {\bf
k}\vert\gg \Lambda$, by
\begin{equation}
\frac{d\sigma^{dA\rightarrow gX}}{d\eta d^{2}{\bf k} d^{2}{\bf
b}}=\frac{2\alpha_{s}}{ C_{F}}\frac{\varphi_{A}({\bf k
},Y+\eta,{\bf b })}{{\bf k }^{2}}\int d^{2}{\bf B }
\int_{\Lambda}^{\vert {\bf k}\vert} \varphi_{d}({\bf q
},Y-\eta\simeq0,{\bf B })d^{2}{\bf q }.
\end{equation}
Using the relation,

\begin{equation}
xG(x_{d},{\bf k }^{2})=\int d^{2}{\bf B } \int_{\Lambda}^{\vert {\bf k}\vert }
\varphi_{d}({\bf q },Y-\eta \simeq 0,{\bf B })\frac{d^{2}{\bf q
}}{\pi},
\end{equation}
we end up with
\begin{equation}
\frac{d\sigma^{dA\rightarrow gX}}{d\eta d^{2}{\bf k} d^{2}{\bf
b}}=\frac{\alpha_{s}(2\pi)}{ C_{F}}\frac{\varphi_{A}({\bf
k},Y+\eta,{\bf b})}{{\bf k}^{2}}xG(x_{d},{\bf k}^{2}).
\end{equation}
Notice the collinear factorization of the gluon distribution in
the proton. For hadron production one has to add the valence quark
contribution which may be written \cite{JMG}
\begin{equation}
\frac{d\sigma^{dA\rightarrow qX}}{d\eta d^{2}{\bf k} d^{2}{\bf
b}}=\frac{\alpha_{s}(2\pi)}{ N_{c}}\frac{\varphi_{A}({\bf k}
,Y+\eta,{\bf b})}{{\bf k}^{2}}xq_{V}(x_{d},{\bf k}^{2}).
\end{equation}
Convoluting with  the fragmentation functions we get
\begin{equation}
\frac{d\sigma^{dA\rightarrow hX}}{d\eta d^{2}{\bf k} d^{2}{\bf
b}}=\frac{\alpha_{s}(2\pi)}{ C_{F}}\sum_{i=g,u,d}\int_{z_0}^1
dz\frac{\varphi_{A}({\bf k}/z,Y+\eta+\ln z,b)}{{\bf k}^{2}}[
f_{i}(x_{d}/z,{\bf k}^{2}/z^{2})D_{h/i}(z,{\bf k}^2)],
\end{equation}
where $ f_{u,d}(x,{\bf k}^{2})=(C_{F}/N_{c})xq_{u,d}(x,{\bf
k}^{2})$ and $f_{g}(x,{\bf k}^{2})=xG(x,{\bf k}^{2})$ are the
parton distributions inside the proton; $D_{h/i}(z,{\bf k})$ are
F.F's of the parton $i$ into hadron $h$, and
$z_0=(k_{\bot}/\sqrt{s})e^{\eta}$. Notice that this formula is
identical to eq. (22) of \cite{dum} in the leading twist
approximation. At forward rapidity, the longitudinal momentum
fraction $x_d$ carried by the parton inside the deuteron is of
order of one. This implies that the dynamics of the deuteron is
completely dominated by valence quarks. For numerical calculations
we use the GRV LO parton distribution functions inside the proton
from \cite{grv}, assuming that there is no significant difference
between the proton and the neutron for charged hadron production.
\subsubsection{The unintegrated gluon distribution}
\label{121}
  In the physics of saturation the relevant observable is
the forward scattering amplitude of a quark-antiquark dipole off a
target (a nucleus in our case). It enters several processes at
high energy like DIS, photoproduction and hadron-hadron
scattering. Its quantum evolution has recently been the object of
many studies \cite{KKT1,KT,BK,munp,munp2,MuT,IIMc}. The Balitsky-Kovchegov (BK) equation \cite{BK} valid in the large
$N_{c}$ limit and in the mean field approximation provides a
tool to study the rapidity behavior of the gluon distribution in a large momentum range including the saturation region  where the
effects of gluon recombination, taken into account in the
non-linear term, become important, instead of the BFKL equation
\cite{BFKL} which contains only gluon splitting (linear term)
yielding  a power growth of the gluon distribution with respect to
the center of mass energy violating unitarity and the Froissart
bound. In this work, as a simple test of the theory, we focus on the region above the saturation scale. The BK equation reads in momentum space
\begin{equation}
\partial_{y}\tilde{N}({\bf k},y)=\frac{\alpha
N_{c}}{\pi}[\chi(-\frac{\partial}{\partial\ln{\bf
k}^2})\tilde{N}({\bf k},y)-\tilde{N}^{2}({\bf k},y)],
\end{equation}
where $\chi(\gamma)=2\psi(1)-\psi(\gamma)-\psi(1-\gamma)$ is the
BFKL kernel, and

\begin{equation}\label{NN}
\tilde{N}({\bf k},{\bf b},y)=\int \frac{d^{2}{\bf z}}{{\bf
z}^{2}}N({\bf z},{\bf b},y)e^{i{\bf z}{\bf k}}.
\end{equation}
At large $y$, a solution has been found for the BK equation above and not too far from
the saturation scale in terms of travelling waves. It has a
geometric scaling behavior in the variable $L=\ln ({\bf
k}^{2}/Q_{s}^{2}({\bf b},y))$ when $y$ goes to infinity
\cite{munp,MuT}

\begin{equation}\label{BKsol}
\tilde{N}(L,Y)=CL\exp[-\gamma_{s}L-\beta (y) L^{2}].
\end{equation}
C is an undetermined constant irrelevant in the present analysis
of $R_{dA}$ and $R_{CP}$, $\gamma_s\simeq 0.628$ is the anomalous
dimension of the BFKL dynamics in the geometric scaling region
\cite{MuT,IIMc}: $Q_s^2 \lesssim k_{\bot}^2 \ll
Q_s^2(Y)\exp{(1/\beta)} $ and
$\beta\equiv\beta(y)=(2\bar{\alpha}\chi''(\gamma_{s})y)^{-1}$. We
used the recent fit to the HERA data performed in \cite{IIM} where
$\beta=(2\lambda\kappa y)^{-1}$, $\lambda=0.25$, $\kappa=9.9$ and
$Q_{s}^2(y,0)=3/2Q^2_{s.min.bias}(y)=3/2A^{1/3}(x_0/x)^{\lambda}$
GeV$^2$ with $x_0=0,67.10^{-4}$ and $x=e^{-y}$ given here by
$x=k_{\bot}/\sqrt{s}e^{-\eta}$. There is no straightforward way to
link this asymptotic form to the one at mid-rapidity: this implies
that there is an overall constant coming from $Q_s$ which is
negligible at very large rapidity and far from the saturation
scale. We expect this constant to play a significant role at RHIC
energies. Then, one can fix $Q_{s}$ as shown above, and put all
the freedom into an additional constant ($L\rightarrow L+L_{0}$).
Thus,
\begin{equation}\label{Ntilde}
\tilde{N}(L,y)=C'(L+L_{0})\exp[-(\gamma_{s}+2\beta L_{0})L-\beta
L^{2}].
\end{equation}
Making the same approximations as in the BK equation, one can
write an expression for $N_{G}$ \cite{KKT1}
\begin{equation}
N_{G}({\bf z},y)=\frac{2C_{F}}{N_{c}}(2N({\bf z},y)-N^{2}({\bf
z},y)).
\end{equation}
In the leading twist approximation where one can neglect
non-linear terms, we find
\begin{equation}
N_{G}({\bf z},y)\simeq 2N({\bf z},y).
\end{equation}
Both distributions $\tilde{N}$ and $\varphi_{A}$ are linked to the
dipole scattering amplitude in (\ref{NN}) and (\ref{Nphi}). One
can thus eliminate the latter yielding
\begin{equation}
\varphi_{A}(L,y)=\frac{4N_{c}}{\alpha_{s}(2\pi)^{3}}\frac{d^2}{dL^2}\tilde{N}(L,y).
\end{equation}
At very large $y$ one can neglect the term which breaks scaling,
namely $\beta \ll 1$. So that the last expression reduces to the
simple form exhibiting an exact scaling behavior
\begin{equation}\label{ESphi}
\varphi_{A}(L,y)\varpropto(L+L_{0}-\frac{2}{\gamma_{s}})\exp[-\gamma_{s}L].
\end{equation}
We have fixed $L_{0}$ such that $\varphi_{A}$ has a maximum when
${\bf k}^2=Q_{s}^{2}({\bf b},y)$ \cite{MuT,BMS} corresponding to
$L_{0}=3/\gamma_{s}$. This is the only free parameter of our
calculation, it exhibits the uncertainty in the value of $Q_s$. It
turns out that $R_{CP}$ is very sensitive to variations of $L_{0}$
at energies of RHIC. For the numerical study we choose three
different expressions for $\varphi_A$,
selecting various specific terms in (\ref{Ntilde}) :\\\\
i) The BFKL saturation-inspired form \cite{IIMc}, which violates
scaling, derived from

\begin{equation}\label{BFKL}
\tilde{N}(L,y)_{BFKL}\varpropto\exp[-\gamma_{s}L-\beta L^{2}].
\end{equation}
ii) The BK exact-scaling form (\ref{ESphi}). \\
iii) Finally the full expression derived from (\ref{Ntilde}).\\
\\

\begin{figure}[hbtp]
\centering
 \begin{tabular}{c c c }

\qquad\includegraphics[width=4cm]{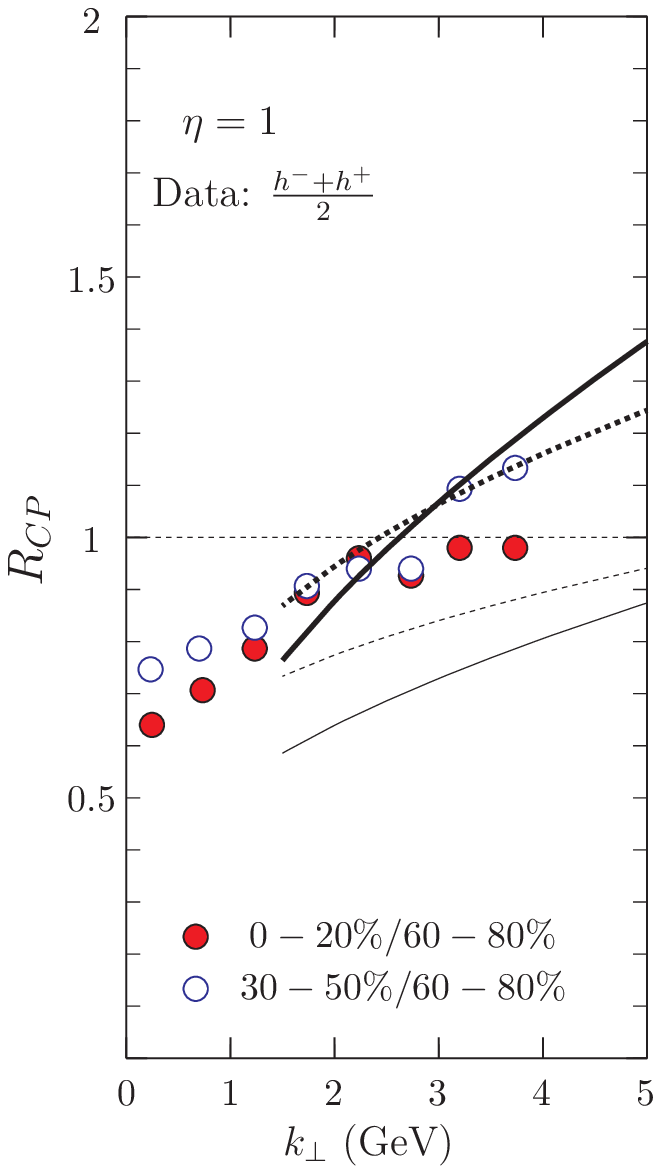} &\qquad   \includegraphics[width=4cm]{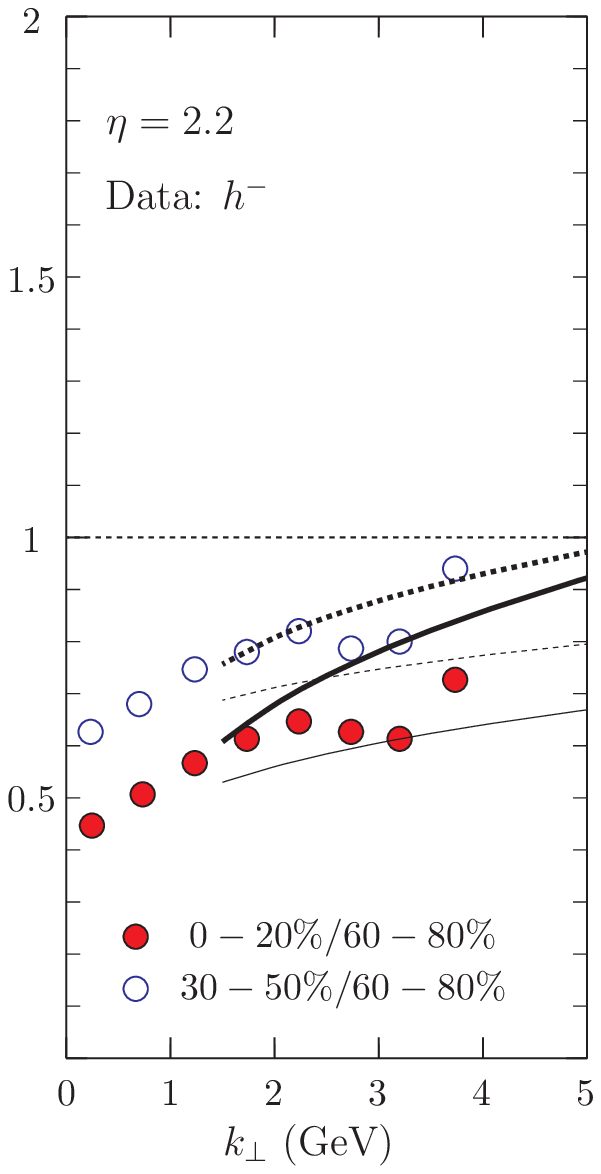} &\qquad   \includegraphics[width=4cm]{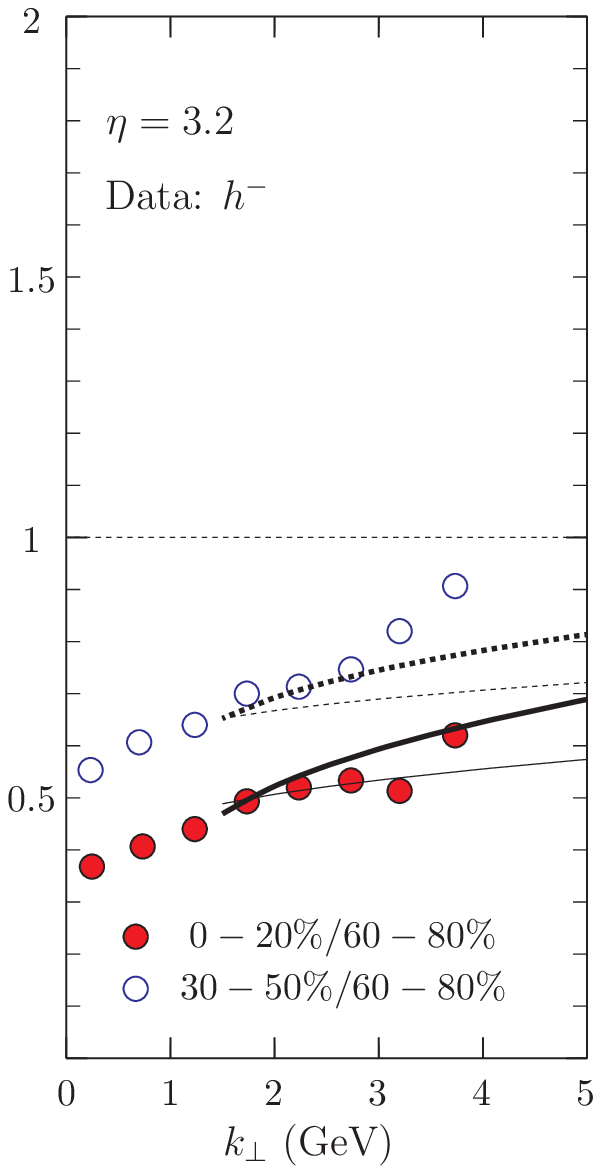}\\
      \qquad (a)& \qquad(b)& \qquad(c)
\end{tabular} \caption{
$R_{CP}$ for the BK parametrization (thick lines) and the
BFKL + saturation form (thin lines) at different rapidities $\eta=1,
2.2$ and $3.2$. Full lines correspond to central over peripheral
collisions (full experimental dots). Dashed lines correspond to
semi-central over peripheral collisions (empty experimental dots
). Data from \cite{BRAHMS1}.} \label{fig4}
\end{figure}
\begin{figure}[hbtp]
\centering
 {\includegraphics[width=4cm]{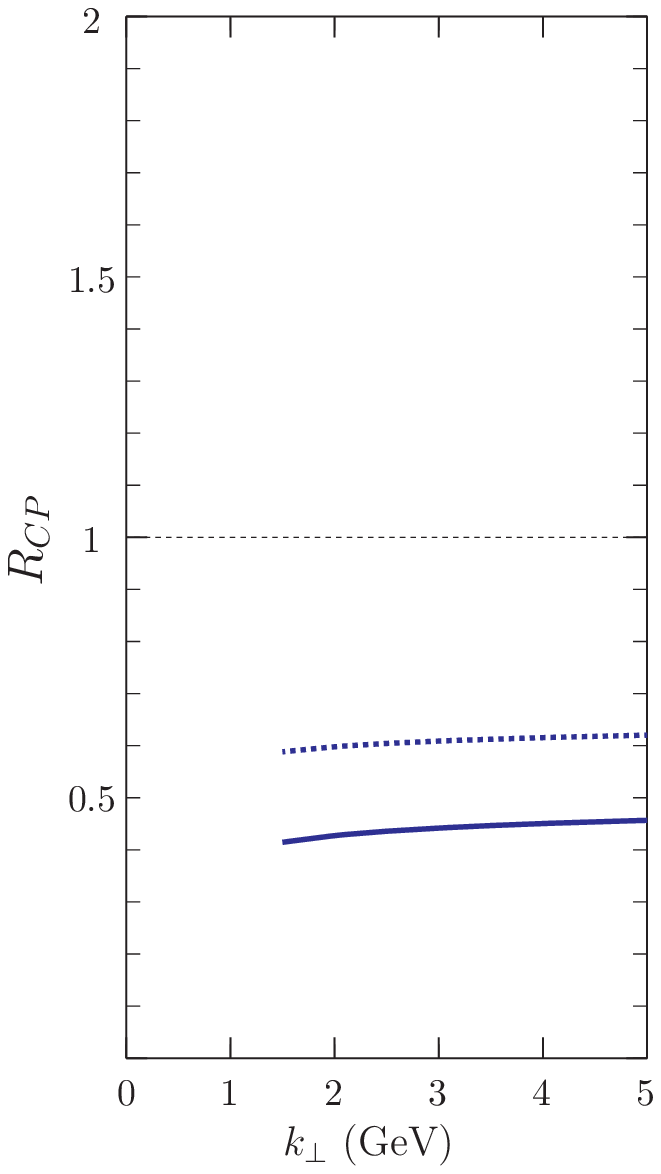}}

\caption{$R_{CP}$ for the exact scaling BK form at $\eta=3.2$.
Full line : central/peripheral, dashed line:
semi-central/peripheral. } \label{fig5}
\end{figure}

In Fig. \ref{fig4} we show the comparison between the BFKL +
saturation form (\ref{BFKL}) and the BK parametrization
(\ref{Ntilde}). We expect this comparison to be valid at large
enough $k_{\bot}$. The agreement with data is quite good for the
latter. With decreasing rapidity we would expect our formula to
break down, nevertheless the global features of the data are
reproduced even at $\eta=1$. The exact scaling form, shown in Fig.
\ref{fig5}, is a slowly varying function of $\eta$ and is too low
to describe the data. However, if the picture is right, for
increased rapidity ($\eta\simeq 5$ or $6$), data points should
match that shape. The fact that the saturation model has a
semi-quantitative agreement with data for $R_{CP}$ is essentially
due to the anomalous dimension since an approximate form is (when
$k_{\bot}\gtrsim Q_s$)
\begin{equation}
R_{CP}\simeq
\frac{N^{P}_{coll}}{N^{C}_{coll}}(\frac{N^C_{part}}{N^P_{part}})^{\gamma_{eff}}\simeq
(\frac{N^C_{part}}{N^P_{part}})^{\gamma_{eff}-1}.
\end{equation}
 At forward rapidity
$\gamma_{eff}\simeq\gamma_s+\beta(\eta)\ln(k_{\bot}^2/Q_s^2)$ is a
decreasing function of $\eta$ and an increasing function of
$k_{\bot}$. This allows us to understand the qualitative behavior
shown by data and in particular the inversion of the centrality
dependence compared to mid-rapidity where $R_{CP}\gtrsim 1$
(Cronin enhancement) corresponding to $"\gamma_{eff}"\gtrsim 1$ .
At very large $\eta$ the anomalous dimension stabilizes at
$\gamma_{eff}=\gamma_s$, which could be tested at the LHC.

\section{Conclusion and outlook}
To conclude we may reiterate that in our opinion, the saturation
picture as probed by mid-rapidity data for the nuclear
modification factor at RHIC energies cannot be seen as predictive.
Indeed, the CGC at mid-rapidity is based on leading order pQCD
calculations including multiple eikonal partonic rescatterings
inside the nucleus, giving rise to a saturation scale $Q_s^2$ at
most of order of 2 GeV$^2$ and it turns out that this is not
sufficient to describe RHIC data at mid-rapidity. It may be very
interesting to relate the increase of the saturation scale needed
to explain the dA data to the comparison of experimental
measurements with theory for observables such as jet broadening in
nucleus-nucleus (AA) collisions and also jet quenching, since the
saturation scale
determines their order of magnitude. \\

At forward rapidity, a  saturation inspired framework for quantum
evolution predicts the suppression as observed in data. This is a
leading twist effect driven by the anomalous dimension which is
the main source of the observed behavior. There is good evidence
that the answer to the question: "has saturation been observed at
RHIC ?" is positive and a definitive answer could be provided if
one measures, in a more precise way, the anomalous dimension as
predicted by the theory. To do so, one needs to go to larger
rapidities (energies) at LHC.
\\

As one more step to improve further the status of the present
theoretical description  let us remark that initial state
suppression effects may be also present in AA collisions at large
$\eta$ and large transverse momentum $k_{\bot}$. An interesting
quantity to be measured is the double ratio
$R(\eta_1,\eta_2,k_{\bot})$ with $\eta_1>\eta_2>0$, defined by
\begin{equation}
R(\eta_1,\eta_2,k_{\bot})=\frac{R_{CP}(\eta_1,k_{\bot})}{R_{CP}(\eta_2,k_{\bot})},
\end{equation}
as a function of $k_{\bot}$. For $\eta_1=2.2$ and $\eta_2=0$, this
ratio is measured for AA by the BRAHMS Collaboration \cite{Reta};
there is evidence of suppression becoming more important at
forward
rapidities, as observed in dA measurements.\\

Because of the final state suppression in AA collisions one may
argue that this ratio is bounded by
\begin{equation}\label{eta12}
R(\eta_1,\eta_2,k_{\bot})<R(\eta_1,\eta_2,k_{\bot})_{initial}<1,
\end{equation}
where $R(\eta_1,\eta_2,k_{\bot})_{initial}$ for AA collisions may
be estimated for $R_{CP-initial}$ in a way done for dA collisions
(e.g. by assuming $k_{\bot}-$factorization). We expect it to be
quantitatively similar to the dA case (e.g. Figs. \ref{fig4}(b)
and (c)), which is well understood in the saturation picture for
$\eta_1>\eta_2>0$. The inequality (\ref{eta12}) is based on the
assumption that  $R_{CP}$ in AA collisions may be factorized as
$R_{CP-initial}Q(k_{\bot},\eta)$: the quenching factor takes into
account the (radiative) final state interactions \cite{BaierQ}. It
depends, at fixed $k_{\bot}$ on the pathlength of the parton
propagating in a dense medium, such that $Q$ is decreasing with
increasing pathlength, i.e. it amounts to more quenching.
Assuming, on geometrical grounds, a longer path for $\eta_1$ than
for $\eta_2$, $\eta_1>\eta_2$, implies
$Q(k_{\bot},\eta_1)/Q(k_{\bot},\eta_2)<1$, and (\ref{eta12})
follows.\\\\\\
\begin{Large}
\textbf{Acknowledgements}
\end{Large}
\\\\
We acknowledge helpful comments by A. H. Mueller.

\end{document}